\documentclass[twocolumn,showpacs,preprintnumbers]{revtex4}
\usepackage{amssymb}
\usepackage{amsmath}
\usepackage{graphicx}
\usepackage{dcolumn}
\usepackage{bm}
\usepackage{color}

\begin{document}

\title{Quantum phase transitions in the spin-boson model without the
counterrotating terms}
\author{Yan-Zhi Wang$^{1}$, Shu He$^{2}$, Liwei Duan$^{1}$, and Qing-Hu Chen$%
^{1,3,*}$}

\address{
$^{1}$ Department of Physics and Zhejiang Province Key Laboratory of Quantum Technology and Device, Zhejiang University, Hangzhou 310027, China \\
$^{2}$ Department of Physics and Electronic Engineering,
Sichuan Normal University, Chengdu 610066, China\\
$^{3}$  Collaborative Innovation Center of Advanced Microstructures, Nanjing University, Nanjing 210093, China
 }\date{\today }

\begin{abstract}
We study the spin-boson model without the counterrotating terms by a
numerically exact method based on variational matrix product states.
Surprisingly, the second-order quantum phase transition (QPT) is observed
for the sub-Ohmic bath in the rotating-wave approximations. Moreover,
first-order QPTs can also appear before the critical points. With the
decrease of the bath exponents, these first-order QPTs disappear
successively, while the second-order QPT remains robust. The second-order
QPT is further confirmed by multi-coherent-states variational studies, while
the first-order QPT is corroborated with the exact diagonalization in the
truncated Hilbert space. Extension to the Ohmic bath is also performed, and
many first-order QPTs appear successively in a wide coupling regime, in
contrast to previous findings. The previous pictures for many physical
phenomena for the spin-boson model in the rotating-wave approximation have
to be modified at least at the strong coupling.
\end{abstract}

\maketitle

\section{Introduction}

The spin-boson model describes a qubit (two-level system) coupled to a quantum
environment represented by a continuous bath of bosonic modes. It is a
paradigmatic model in many fields, ranging from quantum optics \cite{Scully}%
to condensed matter physics \cite{Leggett}  to open quantum systems \cite%
{Breuer,weiss1}. The Hamiltonian is given by
\begin{equation}
H_{SB}=\frac{\Delta }{2}\sigma _{z}+\sum_{k}\omega _{k}a_{k}^{\dag
}a_{k}+\sum_{k}g_{k}\left( a_{k}^{\dag }+a_{k}\right) \sigma _{x},
\label{spinboson}
\end{equation}%
where $\sigma _{i}$ ($i=x,y,z$) is the Pauli matrices, $\Delta $ is the
qubit frequency, $a_{k}$ ($a_{k}^{\dag }$) is the bosonic annihilation
(creation) operator which can annihilate (create) a boson with frequency $%
\omega _{k}$, and  $g_{k}$ denotes the coupling strength between the qubit and
the bosonic bath, which is usually characterized by the power-law spectral
density $J(\omega )$,
\begin{equation}
J(\omega )=\pi \sum_{k}g_{k}^{2}\delta (\omega -\omega _{k})=2\pi \alpha
\omega _{c}^{1-s}\omega ^{s}\Theta (\omega _{c}-\omega ),
\end{equation}%
where $\alpha $ are a dimensionless coupling constant, $\omega _{c}$ is the
cutoff frequency, and $\Theta (\omega _{c}-\omega )$ is the Heaviside step
function. The bath exponent $s$ classifies the reservoir into super-Ohmic $%
\left( s>1\right) $, Ohmic $\left( s=1\right) $, and sub-Ohmic $\left(
s<1\right) $ types. In many theoretical studies, due to the
weak coupling strength in the real quantum optical and quantum dissipative
systems, the counterrotating terms involving higher excited states, $%
a_{k}^{\dag }\sigma _{+}$ and $a_{k}\sigma _{-}$, can be neglected, a condition which is the
so-called  rotating-wave approximation (RWA); thus the full Hamiltonian (%
\ref{spinboson}) can be reduced to the following RWA form%
\begin{equation}
H_{SB}^{RWA}=\frac{\Delta }{2}\sigma _{z}+\sum_{k}\omega _{k}a_{k}^{\dag
}a_{k}+\sum_{k}g_{k}\left( a_{k}^{\dag }\sigma _{-}+a_{k}\sigma _{+}\right) .
\label{RWA}
\end{equation}%
It is generally believed that the RWA is a reasonably good approximation because
the counterrotating terms violate energy conservation, leading to virtual
processes, and thus are suppressed.

The total excitation number of the spin-boson model is $\hat{N}%
=\sum_{k}a_{k}^{\dag }a_{k}+\sigma _{+}\sigma _{-}$, where $\sigma _{\pm
}=\left( \sigma _{x}\pm i\sigma _{y}\right) /2$. With (without) the RWA, the
system possesses a $U(1)$ ($Z_{2}$) symmetry, which is characterized by the
action of the operator $\Re \left( \theta \right) =\exp \left( i\theta \hat{N%
}\right) $, where the arbitrary angle $\theta $ corresponds to $U(1)$
symmetry and the special value of $\theta =\pi \ $\  corresponds to $Z_{2}$ (parity)
symmetry. The parity operator $\hat{\Pi}=\Re \left( \pi \right) $ has two
eigenvalues \ $\pm 1$. \

With the advance of modern technology,  various qubit and oscillator
coupling systems can be engineered in many solid-state devices, such as
superconducting circuits \cite{Niemczyk,Yoshihara}, cold atoms \cite{Dimer},
and trapped ions \cite{Cirac}. Recently, the spin-boson model has been
realized by the ultrastrong coupling of a superconducting flux qubit to an
open one-dimensional (1D) transmission line ~\cite{Forn2}. The counterrotating terms can be
strongly suppressed in some proposed schemes \cite%
{AnistropicDicke,Keeling,Fanheng}. In some systems, the anisotropy appears
quite naturally, because they are controlled by different input parameters~
\cite{Grimsmo1}. On the theoretical side, the anisotropic models where the
rotating and counterrotating terms are different have attracted
considerable attentions. Rich quantum phase transitions (QPTs) in the
anisotropic Dicke models including the Tavis-Cummings model \cite{TC}
without counterrotating terms have been reported \cite{AnistropicDicke} in
the thermodynamic limit, i.e., infinite qubits. It was found in~\cite%
{JC_Plenio} that the Jaynes-Cummings (JC) model ~\cite{JC}, the quantum Rabi
model in the RWA, could also undergo the second-order QPT in the extreme
model parameter limit, $\Delta /\omega \rightarrow \infty $ where $\Delta $
and $\omega $ are the frequencies of qubit and cavity. It was demonstrated
that the ratio of frequencies $\Delta /\omega $ plays the same role as the
qubits number, and the second-order QPT can be identified by  finite
frequency scaling analysis.

Since both the Dicke and Rabi models in the RWA undergo the second-order
QPT, can the spin-boson model in the RWA  display the
second-order QPT? Are some extreme conditions for model parameters required
to realize the second-order QPT? As we know from the literature, the spin-boson Hamiltonian with the RWA is usually treated in 
sub-space with  fixed total excitations $N$. In this way the second-order
QPT is absolutely excluded. The constrained condition for the given
sub-space should be relaxed to detect more physical phenomena.

In the sub-Ohmic spin-boson model without the RWA, the second-order QPT from the
delocalized phase, where spin has the equal probability in the two states, to
localized phase, in which  spin prefers to stay in one of the two states, has
been studied extensively \cite{Bulla,QMC,ED, Zhang,guo2012critical,Frenzel,CRduan}.
Many advanced numerical approaches in  quantum many-particle physics have
been applied and extended to this model, such as the numerical
renormalization group~\cite{Bulla}, quantum Monte Carlo simulations~\cite%
{QMC}, sparse polynomial space approach~\cite{ED}, exact diagonalization in
terms of shift bosons \cite{Zhang}, and various matrix product state
approaches \cite{guo2012critical,Frenzel}. Some analytical approaches based
on the polaronic unitary transformation, also known as the Silbey-Harris ansatz ~%
\cite{Silbey}, have been also developed for this model \cite%
{zheng,Chin,mD1,multi_CS2,yzhao,Blunden,ZH,cao,HeShu1}. The single-coherent-states ansatz~\cite{Silbey} was improved by simply adding other coherent
states on an equal footing~\cite{mD1} and by superpositions of two
degenerate single coherent states~\cite{ZH}, which are generally termed as the
multi-coherent-states (MCS) ansatz. Actually, the MCS in the single-mode
model was proposed much earlier by Ren and Chen~\cite{tcs}.

In this paper, we will extend the variational matrix product state (VMPS)
approach ~\cite{guo2012critical} to study the spin-boson model in the RWA
for all values of the bath exponents. The MCS variational approach and exact
diagonalization within truncated Hilbert space are also employed to provide
independent checks in different regimes. The paper is organized as follows.
In Sec. II, we briefly introduce the generalized spin-boson model. Some
methodologies including the VMPS, the MCS variational approaches, and exact
diagonalization in truncated Hilbert space are described. The rich phase
transitions revealed by the VMPS method are presented in Sec.  III, where
the MCS variational approaches and the exact diagonalization are also
applied to provide further evidence. The quantum criticality based on VMPS
studies is also analyzed. Finally, conclusions are drawn in Sec. IV.

\section{Generalized Model Hamiltonian and methodologies}

Based on Hamiltonian (\ref{spinboson}), the generalized spin-boson
Hamiltonian can be written as ($\hbar =1$)
\begin{widetext}
\begin{equation}
\hat{H_{SB}} =\frac{\Delta }{2}\sigma _{z}+\frac{\epsilon }{2}\sigma
_{x}+\sum_{k}\omega _{k}a_{k}^{\dag }a_{k}
+\frac{1+\lambda }{2}\sum_{k}g_{k}\left( a_{k}^{\dag }\sigma
_{-}+a_{k}\sigma _{+}\right) +\frac{1-\lambda }{2}\sum_{k}g_{k}\left(
a_{k}\sigma _{-}+a_{k}^{\dag }\sigma _{+}\right) ,  \label{Hamiltonian1}
\end{equation}
\end{widetext}
where $\epsilon $ is the energy bias applied in a two-level system ($\epsilon
=0$ except special statements) and $\frac{1+\lambda }{2}\ $and $\frac{1-\lambda
}{2}\ $ are the weights of the rotating-wave and counterrotating terms,
respectively. In this sense $\lambda $ is the anisotropy constant of this
model. Obviously, $\lambda =1$ ($\lambda =0$) corresponds to the spin-boson
model with (without) RWA.

For later use, Hamiltonian (\ref{Hamiltonian1}) can be rewritten as
\begin{eqnarray}
\hat{H} &=&\frac{\Delta }{2}\sigma _{z}+\frac{\epsilon }{2}\sigma
_{x}+\sum_{k}\omega _{k}a_{k}^{\dag }a_{k}+\frac{1}{2}\sum_{k}g_{k}\left(
a_{k}^{\dag }+a_{k}\right) \sigma _{x}  \notag \\
&&+\frac{\lambda }{2}\sum_{k}g_{k}\left( a_{k}-a_{k}^{\dag }\right) i\sigma
_{y}.  \label{Hamiltonian2}
\end{eqnarray}%
In the following, three methods are introduced to study this generalized
model.

\textsl{VMPS approach.-}. As is well known, the VMPS approach works efficiently
in one-dimensional chain models ~\cite{MPS_intro1,MPS_intro2}. To apply VMPS
in the spin-boson model, we therefore transform the model into a 1D chain
model. We first perform the logarithmic discretization of the spectral
density of the continuum bath~\cite{Bulla} with discretization parameter $%
\Lambda >1$; then, by using orthogonal polynomials $b_{n}^{\dag
}=\sum_{k}U_{nk}a_{k}^{\dag }$ ($b_{n}=\sum_{k}U_{nk}a_{k}$) as described in
Ref.~\cite{Plenio_Chin}, the spin-boson models can be mapped into the
representation of a one-dimensional semi-infinite chain with
nearest-neighbor interaction ~\cite{Friend}. Thus, Hamiltonian (\ref%
{Hamiltonian2}) can be written as:
\begin{eqnarray}
H_{\text{chain}} &=&\frac{\Delta }{2}\sigma _{z}+\frac{\epsilon }{2}\sigma
_{x}+\frac{c_{0}}{2}(b_{0}+b_{0}^{\dag })\sigma _{x}+\lambda \frac{c_{0}}{2}%
(b_{0}-b_{0}^{\dag })i\sigma _{y}  \notag \\
&&+\sum_{n=0}^{L-2}[\epsilon _{n}b_{n}^{\dag }b_{n}+t_{n}(b_{n}^{\dag
}b_{n+1}+b_{n+1}^{\dag }b_{n})],  \label{Hamitrans}
\end{eqnarray}%
where $b_{n}^{\dag }$($b_{n}$) is the creation (annihilation) operator for a new
set of boson modes in a transformed representation with $\epsilon _{n}$
describing frequency on chain site $n$, $t_{n}$ describing the
nearest-neighbor hopping parameter, and $c_{0}$ describing the effective
coupling strength between the spin and the new effective bath. All of the
parameters mentioned above, such as $t_{n},\epsilon _{n}$ and $c_{0}$, are
determined by the logarithmic discretization parameter $\Lambda $, the
cutoff frequency $\omega _{c}$, and a specific form of the spectral
density, which are expressed below{\
\begin{eqnarray*}
c_{0} &=&\sqrt{\int_{0}^{\omega _{c}}\frac{J\left( \omega \right)}{\pi}
d\omega} , \\
\epsilon _{n} &=&\xi _{s}\left( A_{n}+C_{n}\right) , \\
t_{n} &=&-\xi _{s}\left( \frac{N_{n+1}}{N_{n}}\right) A_{n},
\end{eqnarray*}%
where
\begin{eqnarray*}
\xi _{s} &=&\frac{s+1}{s+2}\frac{1-\Lambda ^{-\left( s+2\right) }}{1-\Lambda
^{-\left( s+1\right) }}\omega _{c}, \\
A_{n} &=&\Lambda ^{-j}\frac{\left( 1-\Lambda ^{-\left( j+1+s\right) }\right)
^{2}}{\left( 1-\Lambda ^{-\left( 2j+1+s\right) }\right) \left( 1-\Lambda
^{-\left( 2j+2+s\right) }\right) }, \\
C_{n} &=&\Lambda ^{-\left( j+s\right) }\frac{\left( 1-\Lambda ^{-j}\right)
^{2}}{\left( 1-\Lambda ^{-\left( 2j+s\right) }\right) \left( 1-\Lambda
^{-\left( 2j+1+s\right) }\right) }, \\
N_{n}^{2} &=&\frac{\Lambda ^{-n\left( 1+s\right) }\left( \Lambda
^{-1}:\Lambda ^{-1}\right) _{n}^{2}}{\left( \Lambda ^{-\left( s+1\right)
}:\Lambda ^{-1}\right) _{n}^{2}\left( 1-\Lambda ^{-\left( 2n+1+s\right)
}\right) },
\end{eqnarray*}%
with }%
\begin{equation*}
\left( a:q\right) _{n}=\left( 1-a\right) \left( 1-aq\right) ...\left(
1-aq^{n-1}\right)
\end{equation*}%
For details, one may refer to Ref.~\cite{Plenio_Chin}.

We now briefly introduce the VMPS approach~\cite{VMPS1,VMPS2,MPS_intro1}.
For the transformed spin-boson model of a 1D chain with $L$ sites, the ground-state wave function of Hamiltonian (\ref{Hamitrans}) can be depicted as
\begin{equation}
\left\vert \psi \right\rangle =\sum_{\{N_{n}\}=1}^{d_{n}}M\left[ N_{1}\right]
...M\left[ N_{L}\right] \left\vert N_{1},...,N_{L}\right\rangle ,
\label{MPSWF}
\end{equation}%
where $N_{n}$ is the physical dimension of each site $n$ with truncation $%
d_{n}$, and $D_{n}$ is the bond dimension for matrix $M$ with the open
boundary condition, bounding the maximal entanglement in each subspace. $M$
on each site is optimized through sweeping the 1D chain iteratively, where
accuracy of numerical results is determined by values of $d_{n}$ and $D_{n}$%
. In order to deal with the spin-boson model near the quantum critical region
effectively, we apply an optimized boson basis through an additional
isometric map with $d_{opt}\ll d_{n}$ like in Refs.~\cite%
{guo2012critical,Friend}. In this way, we can effectively reduce the local boson
basis and improve the maximum  boson number in the quantum critical region.

We have recovered all the results in Ref. \cite{guo2012critical} and
confirmed the classical mean-field behavior for $s<1/2$ of the sub-Ohmic
spin-boson model. In the present paper, we will extend it to study the RWA
spin-boson model, mainly focusing on phase transitions in the sub-Ohmic baths,
namely $\lambda =1$, $0<s<1$. For the data presented below, we typically
choose the model parameters as $\Delta =0.1$, $\omega _{c}=1$, and $\epsilon =0$%
, if there are no special statements. Some parameters of orthogonal
polynomials transformation and VMPS are the same as those in Ref.~\cite%
{guo2012critical}, e.g. \ the logarithmic discretization parameter $\Lambda
=2$, the length of the semi-infinite chain $L=50$, and optimized truncation numbers $%
d_{opt}=12$. In addition, we adjust the bond dimension to achieve better
convergence of the results. In this paper, we choose $D_{max}=20,40$ for $%
s=0.3,0.7$, respectively, which is sufficient to obtain the converged
results, as demonstrated in Appendix A in detail.

\textsl{Exact diagonalization in truncated Hilbert space.-}. It is well
known that the spin-boson model with the RWA possesses  $U(1)$ symmetry,
and the total excitation number $\hat{N}$ is conserved because $\left[ \hat{N},H%
\right] =0$. It has been reported in Ref. \cite{Tong} that the total excitation
number $N$\quad in the ground state of the RWA spin-boson model jumps from $%
0 $ to $1$ at a critical coupling strength. This instability can be also
called the first-order phase transition, because the first derivative of the
ground-state energy is discontinuous. Due to the possible sequence of
instabilities with the coupling strength, we can truncate the Hilbert space
up to a finite $N$ excitation number. To this end, we first separate the
Hilbert space into several subspaces with different excitation numbers $%
l=0,1,2...N$. The wave function in $l$-subspace $\left\vert \psi
_{l}\right\rangle $ can be written explicitly with $l$ excitations. E. g.
the wavefunctions for $l=0,1,2$ and $3$-subspace are listed in the following
\begin{eqnarray*}
\left\vert \psi _{0}\right\rangle &=&\left\vert 0\right\rangle \left\vert
\downarrow \right\rangle , \\
\left\vert \psi _{1}\right\rangle &=&c\left\vert 0\right\rangle \left\vert
\uparrow \right\rangle +\sum_{k}d_{k}{a_{k}}^{\dag }\left\vert
0\right\rangle \left\vert \downarrow \right\rangle , \\
\left\vert \psi _{2}\right\rangle &=&\sum_{k}e_{k}{a_{k}}^{\dag }\left\vert
0\right\rangle \left\vert \uparrow \right\rangle +\sum_{kk^{^{\prime
}}}f_{kk^{^{\prime }}}{a_{k}}^{\dag }{a_{k^{\prime }}}^{\dag }\left\vert
0\right\rangle \left\vert \downarrow \right\rangle , \\
\left\vert \psi _{3}\right\rangle &=&\sum_{kk^{^{\prime }}}p_{kk^{^{\prime
}}}{a_{k}}^{\dag }{a_{k^{\prime }}}^{\dag }\left\vert 0\right\rangle
\left\vert \uparrow \right\rangle \\
&&+\sum_{kk^{^{\prime }}k^{^{\prime \prime }}}q_{kk^{^{\prime }}k^{^{\prime
\prime }}}{a_{k}}^{\dag }{a_{k^{\prime }}}^{\dag }{a_{k^{\prime \prime }}}%
^{\dag }\left\vert 0\right\rangle \left\vert \downarrow \right\rangle .
\end{eqnarray*}%
Then the wavefunction in the truncated Hilbert space up to $N$ excitations
can be expressed as%
\begin{equation}
\left\vert \psi \right\rangle ^{\leq N}=\sum_{l=0}^{N}\left\vert \psi
_{l}\right\rangle .  \label{ED}
\end{equation}%
For $N=0$, $\left\vert \psi \right\rangle ^{\leqslant 0}=\left\vert
0\right\rangle \left\vert \downarrow \right\rangle $, the ground-state
energy is $E_{0}=-\frac{\Delta }{2}$. However, it is very difficult to
obtain the analytical solution for $\left\vert \psi \right\rangle
^{\leqslant N}$ up to $N>0$ excitations, and numerically exact diagonalizations
are then required to obtain the converged  lowest energy. This approach is called
$N$ED below.

\textsl{MCS ansatz.-}. We also apply the MCS ansatz \cite{tcs,mD1,multi_CS2}
to the spin-boson model in the RWA. To facilitate the variational study and
visualize the symmetry breaking explicitly, we rotate the Hamiltonian (\ref%
{Hamiltonian2}) around the y axis by an angle $\pi /2$ with $\epsilon = 0$,
which gives
\begin{eqnarray}
H^{T} &=&-\frac{\Delta }{2}\sigma _{x}+\sum_{k}\omega _{k}a_{k}^{\dag }a_{k}+%
\frac{1}{2}\sum_{k}g_{k}\left( a_{k}^{\dag }+a_{k}\right) \sigma _{z}  \notag
\\
&&+\frac{\lambda }{2}\sum_{k}g_{k}\left( a_{k}-a_{k}^{\dag }\right) i\sigma
_{y}  \label{Hamiltonian3}
\end{eqnarray}%
The trial state $|\psi ^{T}\rangle $ is written in the basis of the spin-up
state $|\uparrow \rangle $ and spin-down state $|\downarrow \rangle $
\begin{equation}
|\psi ^{T}\rangle =\left(
\begin{array}{c}
\sum_{n=1}^{N_{c}}A_{n}\exp \left[ \sum_{k=1}^{L}f_{n,k}\left( a_{k}^{\dag
}-a_{k}\right) \right] |0\rangle \\
\sum_{n=1}^{N_{c}}B_{n}\exp \left[ \sum_{k=1}^{L}h_{n,k}\left( a_{k}^{\dag
}-a_{k}\right) \right] |0\rangle%
\end{array}%
\right) ,  \label{VM_wave}
\end{equation}%
where $A_{n}$\ ($B_{n}$) are related to the occupation probabilities of the
spin-up (spin-down) state in the $n$th coherent state; $N_{c}$ and $L$ are
numbers of coherent states and total bosonic modes,  respectively; and $f_{n,k}$ ($%
h_{n,k}$) represents bosonic displacement of the $n$th coherent state and $k$
th bosonic mode. The symmetric MCS ansatz ($A_{n}=B_{n}$ and $%
f_{n,k}=-g_{n,k}$) can only be applied to the delocalized phase, so one can
easily detect the symmetry breaking.

The energy expectation value can be calculated as follows
\begin{equation}
E=\frac{\langle \psi ^{T}|H^{T}|\psi ^{T}\rangle }{\langle \psi ^{T}|\psi
^{T}\rangle },  \label{ehd}
\end{equation}%
where
\begin{eqnarray*}
\langle \psi ^{T}|H^{T}|\psi ^{T}\rangle
&=&\sum_{m,n}(A_{m}A_{n}F_{m,n}\alpha _{m,n} \\
&&+B_{m}B_{n}G_{m,n}\beta _{m,n}-\gamma _{mn}\Gamma _{m,n}A_{m}B_{n}), \\
\langle \psi ^{T}|\psi ^{T}\rangle &=&\sum_{m,n}\left(
A_{m}A_{n}F_{m,n}+B_{m}B_{n}G_{m,n}\right) ,
\end{eqnarray*}%
with
\begin{eqnarray*}
F_{m,n} &=&\exp \left[ -\frac{1}{2}\sum_{k}\left( f_{m,k}-f_{n,k}\right) ^{2}%
\right] , \\
G_{m,n} &=&\exp \left[ -\frac{1}{2}\sum_{k}\left( h_{m,k}-h_{n,k}\right) ^{2}%
\right] , \\
\Gamma _{m,n} &=&\exp \left[ -\frac{1}{2}\sum_{k}\left(
f_{m,k}-h_{n,k}\right) ^{2}\right] , \\
\alpha _{m,n} &=&\sum_{k}\left[ \omega _{k}f_{m,k}f_{n,k}+\frac{g_{k}}{2}%
\left( f_{m,k}+f_{n,k}\right) \right] , \\
\beta _{m,n} &=&\sum_{k}\left[ \omega _{k}h_{m,k}h_{n,k}-\frac{g_{k}}{2}%
\left( h_{m,k}+h_{n,k}\right) \right] , \\
\gamma _{m,n} &=&\left[ \Delta +\lambda \sum_{k}g_{k}\left(
f_{m,k}-h_{n,k}\right) \right] .
\end{eqnarray*}

Minimizing the energy expectation value with respect to variational parameters
gives the following self-consistent equations
\begin{equation*}
\frac{\partial E}{\partial A_{n}}=\frac{\partial E}{\partial B_{n}}=\frac{%
\partial E}{\partial f_{ij}}=\frac{\partial E}{\partial h_{ij}}=0
\end{equation*}%
which leads to
\begin{eqnarray}
&&\sum_{n}\left( 2A_{n}F_{i,n}\left( \alpha _{i,n}-E\right) -\Gamma
_{i,n}B_{n}\gamma _{i,n}\right) =0,  \label{aa} \\
&&\sum_{n}\left( 2B_{n}G_{i,n}\left( \beta _{i,n}-E\right) -\Gamma
_{n,i}A_{n}\gamma _{n,i}\right) =0,  \label{bb} \\
&&\sum_{n}\{-\Gamma _{i,n}B_{n}\left( h_{n,j}\gamma _{i,n}+\lambda
g_{j}\right)  \notag \\
&&+A_{n}F_{i,n}\left[ 2\left( \alpha _{i,n}+\omega _{j}-E\right)
f_{n,j}+g_{j}\right] \}=0,  \label{ff} \\
&&\sum_{n}\{-\Gamma _{n,i}A_{n}\left( f_{n,j}\gamma _{n,i}-\lambda
g_{j}\right)  \notag \\
&&+B_{n}G_{i,n}\left[ 2\left( \beta _{i,n}+\omega _{j}-E\right) h_{n,j}-g_{j}%
\right] \}=0.  \label{hh}
\end{eqnarray}

In practice, these parameters can be obtained by solving the coupled
equations self-consistently, which in turn give the ground-state energy and
wave function. The ground state with zero excitation $\left\vert
0\right\rangle \left\vert \downarrow \right\rangle $ to Hamiltonian (\ref%
{Hamiltonian2}) can be contained in the MCS wave function to Hamiltonian (\ref%
{Hamiltonian3}) by setting the constrained coefficients: $A_{n}=B_{n}$ and$\
f_{n,k}=h_{n,k}=0$\ for all $n$. However the state with nonzero total
excitations cannot be included in the \ MCS ansatz due to the finite number
of coherent states. It has \ been demonstrated that this wave function can
describe the localized phase of the spin-boson model \cite{Blunden}. The
number of the coherent states in the practical calculations in this paper is
$N_{c}=6$, which is sufficient to judge the existence of the second-order
QPT.

For the three approaches described above, discretization of the energy spectrum
of the continuum bath should be performed at the very beginning in the
practical calculations. The same logarithmic discretization is taken for
different approaches if comparison is made in the data presented below.

\section{Results and discussions}

We first describe our main results by VMPS method to the sub-Ohmic $%
\left(0<s<1\right )$ spin-boson model with the RWA described by Eq. (\ref%
{Hamiltonian2}) for $\lambda =1$. We observe a second-order QPT in the
spin-boson model under RWA, which was unnoticed in literature, to the best
of our knowledge. This surprising observation is further confirmed by the
MCS variational approach. The symmetry breaking was unambiguously found
above the critical point in this wave function based approach. The $N$ED
results are also given, which should be exact at  weak coupling and can
be regarded as a benchmark in this regime. Besides the second-order QPT, we
also find a few first-order QPTs before the critical point of the
second-order QPT for large bath exponent $s$. As $s$ decreases, the
first-order QPTs disappear successively, but the second-order phase
transition remains robust no matter how small $s$ is. We will discuss those
phenomena based on various numerical calculations in the following
subsections.

\subsection{Magnetization $\left\vert \left\langle \protect\sigma %
_{x}\right\rangle \right\vert $}

\begin{figure}[tph]
\includegraphics[scale=0.4]{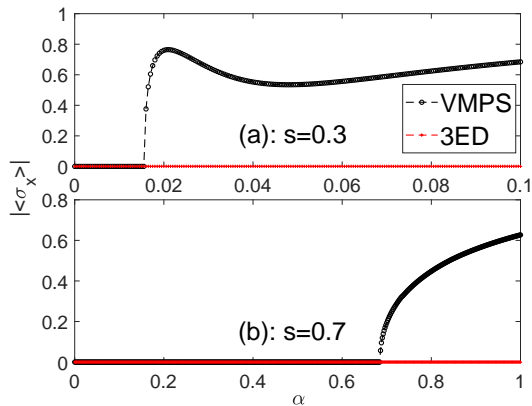}
\caption{ (Color online) Magnetization $\left\vert \left\langle \protect%
\sigma _{x}\right\rangle \right\vert $ as a function of $\protect\alpha $ in
the ground state for (a) $s=0.3$ and (b) $0.7$. Black lines with circles
denote the VMPS results and the red dashed lines denote the $N=3$-ED ones. $\protect%
\lambda =1$, $\Delta =0.1$, $\protect\omega _{c}=1$, $\protect\epsilon =0$, $%
\Lambda =2$, $L=50$, $d_{opt}=12$, and $D=20,40$ for $s=0.3,0.7$,
respectively. }
\label{Magnetization}
\end{figure}
In terms of Hamiltonian (\ref{Hamiltonian2}), magnetization $\left\vert
\left\langle \sigma _{x}\right\rangle \right\vert $ can be regarded as the
order parameter in this model. In the second-order phase transition, due to
the symmetry breaking, the order parameter changes from zero to nonzero at
the critical points. In Fig.~\ref{Magnetization} we present the VMPS results
for $\left\vert \left\langle \sigma _{x}\right\rangle \right\vert $ as a
function of the coupling strength $\alpha $ for the spin-boson model in the
RWA (i.e.$\ \lambda =1$) for two typical values of $s=0.3$ and $0.7$.
Surprisingly, magnetization changes abruptly from zero to nonzero for both
cases. The critical points of the second-order QPTs are $\alpha
_{c}=0.016,0.685$ for $s=0.3,0.7\ $ respectively.

However, the $N$ED within  excitation numbers up to $N$ shows that the
order parameter $\left\vert \left\langle \sigma _{x}\right\rangle
\right\vert $ remains zero for all coupling strengths. Note that the VMPS
has provided  convincing results for the full spin-boson model \cite%
{guo2012critical}. $N$ED cannot yield  consistent results with the VMPS
ones, because the total excitation number may not conserve with $\alpha $
due to the unexpected symmetry breaking.

In the finite-size Dicke model with the RWA \cite{Buzek,Zhangyy}, the system
usually undergoes the first-order phase transition, i.e., sequence of
instabilities, among the phases within different conserved excitation
numbers as the coupling strength increases. In the limit $\Delta /\omega
\rightarrow \infty $, the second-order quantum phase transitions have been
observed in the quantum Rabi model with the RWA \cite{JC_Plenio}. It has been
also reported that the Dicke model under the RWA displays a second-order QPT
in the thermodynamic limit \cite{AnistropicDicke}. Contrary to the quantum
Rabi (Dicke) model in the RWA, the spin-boson model under the RWA could
undergo the second-order QPT even for a finite value of $\Delta /\omega $ (
one qubit).

{\ }

\subsection{Magnetization $\left\vert \left\langle \protect\sigma %
_{z}\right\rangle \right\vert $ and the ground-state energy}

\begin{figure}[tph]
\centerline{\includegraphics[scale=0.4]{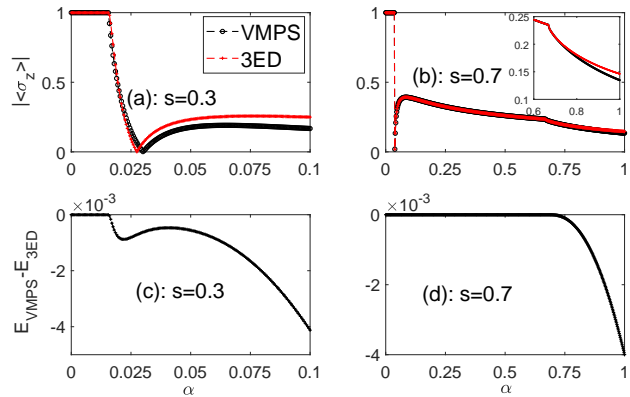}}
\caption{ (Color online) Magnetization $\left\vert \left\langle\protect\sigma%
_{z}\right\rangle\right\vert$ and the ground-state energy difference (lower
panel) by the VMPS and $N=3$ED for $s=0.3$ (left panel) and $0.7$ (right
panel). $\protect\lambda =1$, $\Delta =0.1$, $\protect\omega _{c}=1$, $%
\protect\epsilon =0$, $\Lambda=2$, $L=50$, $d_{opt}=12$, and $D=20,40$ for $%
s=0.3,0.7 $ respectively. The inset in the right-upper panel shows the
enlarged view of the same plot where the kink can be clearly visible.}
\label{GS_quantity}
\end{figure}

{\ In the original spin-boson Hamiltonian, $\sigma _{z}$ describes a
tunneling two-level system ~\cite{zheng,Chin}. The magnetization along the $%
z-$direction }$\left\vert \left\langle \sigma _{z}\right\rangle \right\vert $%
{, simply the order parameter along z direction, is the renormalized factor
of the tunneling amplitude $\Delta $. In this subsection, we examine the
magnetization along the z direction as well as the ground-state energy by
both VMPS and $3$ED, which are exhibited in Fig.~\ref{GS_quantity}. }

Obviously, two observables begin to deviate only after the critical point $%
\alpha _{c}$. Especially we find that the ground-state energies by the VMPS
become lower than those by $3$ED after the critical points, indicating again
the invalidity of ED method at  strong coupling. Practically, we cannot
perform ED with a very large total excitation number due to the huge
Hilbert space. In this paper, our exact diagonalization is only performed up
to $\ N=3$. Of course, if one can really perform  true exact
diagonalization without the limitation of  total excitation numbers, the
true ground state could be also correctly described in the ED method.

{\ It is interesting to note from Figs.~\ref{Magnetization} and \ref%
{GS_quantity} that }$\left\vert {\left\langle \sigma _{x}\right\rangle }%
\right\vert ${\ and }$\left\vert {\left\langle \sigma _{z}\right\rangle }%
\right\vert ${\ exhibit different behaviors with the increase of the
coupling strength. We will discuss this issue based on  parity symmetry
breaking and  total excitation numbers in the  next subsection.}

\subsection{Parity symmetry breaking}

\begin{figure}[tph]
\centerline{\includegraphics[scale=0.4]{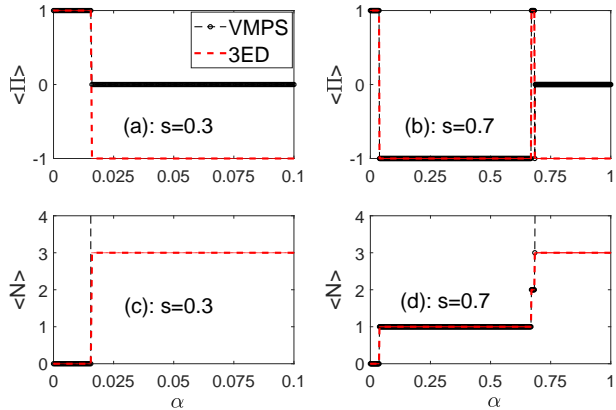}}
\caption{ (Color online) Parity (upper panel) and total excitations (lower
panel) in the ground-state by the three-excitation NED and the present VMPS
for $s=0.3$ and $0.7$. $\protect\lambda =1$, $\Delta =0.1$, $\protect\omega %
_{c}=1$, $\protect\epsilon =0$, $\Lambda=2$, $L=50$, $d_{opt}=12$, and $%
D=20,40$ for $s=0.3,0.7$ respectively. }
\label{parity}
\end{figure}

As stated before, the spin-boson model with (without) the RWA possesses a $%
U(1)$ ($Z_{2}$) symmetry. $U(1)$ is a higher symmetry than $Z_{2}$, so in
the RWA spin-boson model, the system also has $Z_{2}$ symmetry, i.e., 
parity symmetry, like  in the full spin-boson model. In this section, we
study the behavior of the expectation value of the parity $\hat{\Pi}=\exp
\left( i\pi \hat{N}\right) $.

The finite order parameter above the critical points obtained by VMPS in the
previous subsection displays  spontaneous symmetry breaking, while parity
is generally just the criterion to determine whether symmetry is broken. The
upper panel in Fig.~\ref{parity} gives the expectation value of parity $%
\left\langle \hat{\Pi}\right\rangle \ $obtained by both methods for $%
s=0.3,0.7$. Before the critical points $\alpha_c$, both methods yield the
same results for the parity. It is interesting to find that in this regime,
the value of parity for all values of $s<1$ is either $1$ or $-1$, which
corresponds to even or odd parity respectively. It follows that the symmetry
is not broken in this regime. Nevertheless, the average parity $\left\langle
\hat{\Pi}\right\rangle $ becomes zero due to quantum fluctuations above the
critical points. It is not the eigenstate of the parity operator, indicating
 spontaneous parity symmetry breaking. Note also that the parity jumps
for a few times before the critical coupling for $s=0.7$, and remains
unchanged for $s=0.3$.

We also present the average value of the total excitation number $%
\left\langle \hat{N}\right\rangle $ in the ground-state as a function of the
coupling strength$\ \alpha $ in the lower panel of Fig.~\ref{parity}. Below
the critical points, both approaches give the same $\left\langle \hat{N}%
\right\rangle $, which jumps between different plateaus with different
integers, just like in the JC model~\cite{JC_Plenio}. For $s=0.3$, $%
\left\langle \hat{N}\right\rangle=0 $ remains until the second-order
critical point $\alpha_c$, while, for $s=0.7$, $\left\langle \hat{N}%
\right\rangle $ increases from $0$ to $2$ one by one before $\alpha _{c}$.
It follows that the lowest energies in different coupling regime belong to
the energy levels in different $\left\langle \hat{N}\right\rangle $
subspaces, leading to level crossing at some coupling strengths. Thus the
first derivative of the ground-state energy with respect to the coupling
strength must be discontinuous at these coupling strengths, so instability
of $\left\langle \hat{N}\right\rangle $ before $\alpha _{c}$ just
corresponds to the first-order phase transitions. In addition, the jump of $%
\left\langle \hat{N}\right\rangle $ can also account for the back and forth
of the parity between $1 $ and $-1$.

One can also note that the total excitation number$\ \left\langle \hat{N}%
\right\rangle $ by VMPS increases abruptly at the critical coupling $\alpha
_{c}$. The $N$ED approach can only describe the phase with excitation number
less than or equal to $N$. The total excitation number is not limited in
VMPS, so it in principle can describe all phases. The total excitation numbers are
not conserved in the ground state above $\alpha _{c}$, because of the
symmetry breaking. At the critical points, $\left\langle \hat{N}%
\right\rangle $ does not jump to a plateau with the finite larger integer,
different from the first-order phase transitions.

The instability of total excitation number $\left\langle \hat{N}%
\right\rangle $ at the  phase transitions can account for the rich behavior
of $\left\vert \langle \sigma _{z}\rangle \right\vert $ here. For any value
of $s$, {in} the weak-coupling regime, the corresponding ground state is the spin-down state with photonic vacuum for $\left\langle \hat{N}\right\rangle =0$,
i.e., $|\downarrow \rangle |0\rangle $, thus $\langle \sigma _{z}\rangle =-1$%
, as just demonstrated in the upper panel of {Fig. \ref{GS_quantity}} for $%
s=0.3,0.7$. Once $\left\langle \hat{N}\right\rangle \neq 0$ at the first
phase transition no matter whether it is of the first  or the second order,
the spin state in the ground state consists of both spin-up and spin-down
states, which is  drastically different from the spin state only including the
spin-down state at $\left\langle \hat{N}\right\rangle =0$, leading to the
jump of $\left\vert \langle \sigma _{z}\rangle \right\vert $ and $%
\left\vert \langle \sigma _{z}\rangle \right\vert \neq 1$, as exhibited in
Fig. 2. At the later phase transitions between two finite $\left\langle \hat{%
N}\right\rangle \neq 0$, the ground states do not change drastically
because both have spin-up and spin-down states, therefore $\left\vert
\langle \sigma _{z}\rangle \right\vert $ will exhibit kinks, as shown  in
Fig. 2 for $s=0.7$. The drop of $\left\vert \langle \sigma _{z}\rangle \right\vert $ to
zero is an artefact{\ due to} the plot using magnetization, which is the
absolute {\ value} of $\langle \sigma _{z}\rangle $ in our paper. In the
weak-coupling limit, $\langle \sigma _{z}\rangle $ is actually $-1$ as
stated above. In the strong-coupling limit, we always {\ find} $\langle
\sigma _{z}\rangle >0$ in the ground-state, so, {\ in between} $\langle
\sigma _{z}\rangle $ must cross zero, resulting in the drop of $\left\vert
\langle \sigma _{z}\rangle \right\vert $ to zero at some coupling that is
not at any phase transition point. $\left\vert
\langle \sigma _{x}\rangle \right\vert $ is the order parameter of the
second-order QPT, so it only becomes non-zero for couplings larger than the
critical one.

{\ }In short, we find  rich phase transitions in the sub-Ohmic spin-boson
model with the RWA. First, \ the second-order QPT occurs for any finite
model parameters, similar to its counterpart without the RWA. In both the Rabi
model and the Dicke model with the RWA, the second-order QPTs cannot happen for
finite ratio $\Delta /\omega $ \ or finite qubit number. Second, for larger
bath exponents, e.g., $s=$ $0.7$, both the first- and second-order QPTs occur
subsequently with the coupling strength. Several first-order phase
transitions before the critical points indicate a sequence of instabilities.
The first-order phase transition is absent for small bath exponent $s$, such
as for $s=0.3$.

\subsection{Evidence for the second-order QPT by MCS variational studies}

\begin{figure}[tbp]
\centerline{\includegraphics[scale=0.4]{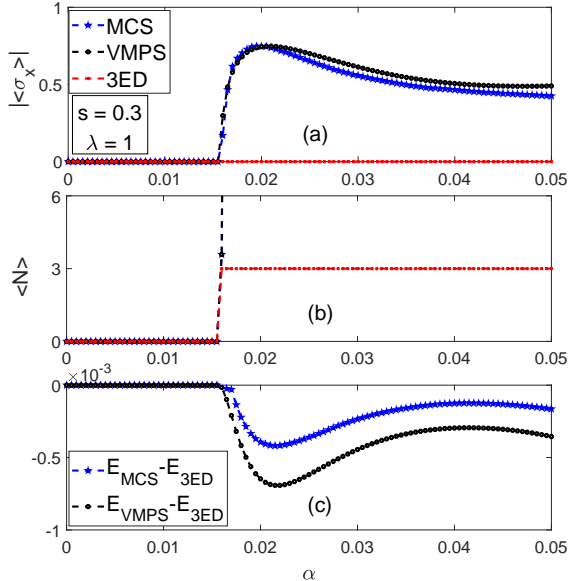}}
\caption{ (Color online) (a) The order parameter $\left\vert \left\langle%
\protect\sigma_{x}\right\rangle\right\vert$.  (b) The total excitation number  $%
\left\langle N \right\rangle$ as a function of the coupling strength within
VMPS, $3$ED, and MCS variational approaches. (c) The difference between the
VMPS (MCS) ground-state energy and that by $3$ED. $s = 0.3$, $\protect%
\lambda =1$, $\Delta =0.1$, $\protect\omega _{c}=1$, $\protect\epsilon =0$, $%
\Lambda=2$, $L=20$, $d_{opt}=12$, $D=20$, $N_c=6$.}
\label{VM}
\end{figure}

Since this model at $s=0.3$ exhibits only one second-order QPT from the zero
excitation ground state, we can employ the MCS variational approach to
provide additional evidence, because the zero excitation can be realized and
the localized phase can be also described in the trial wave function (\ref%
{VM_wave}). In Fig.~\ref{VM}, we list results by  MCS, VMPS, and $N=3$ED
approaches for $s=0.3$. The logarithmic discretization parameter $\Lambda =2$
and $L=20$ bosonic modes are taken for all three approaches here. Note that
the number of bosonic modes here is smaller than those in other figures due
to the computational difficulties in the MCS approach, but it does not
influence the essential results at all. The MCS approach is used here to
account for the existence of the second-order QPT qualitatively, not for the
precise location of the critical points.

Before the critical points, the results by all three methods are the same.
Nonzero order parameter $|\left\langle \sigma _{x}\right\rangle |$ by MCS
approach appears above the critical points, providing   further convincing
evidence of the spontaneous symmetry breaking. The MCS ground-state energies
are lower than those by $N$ED method after the critical point, indicating
that the state with broken symmetry is more stable. The total excitation
number by MCS method increases suddenly above the critical point, because of
no limitation of total excitation number in the coherent state. All these
findings in the MCS variational study provide strong evidence of the
second-order QPT in the spin-boson model with RWA. As found recently by
Blunden-Codd et al.~\cite{Blunden} that a very accurate wave function can
be only obtained by at least $100$ coherent states,  by $N_{c}=6$
coherent states  the MCS results for the order parameter and energy still
slightly deviate from those by VMPS above the critical points.

\subsection{The critical exponent for magnetization}

\begin{figure}[tbp]
\centerline{\includegraphics[scale=0.4]{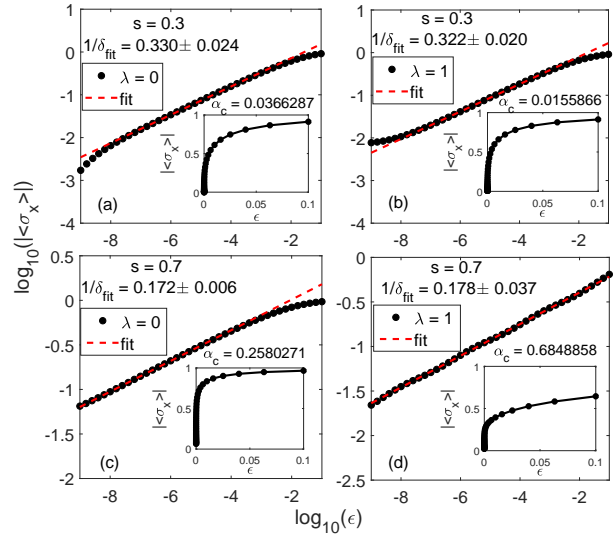}}
\caption{ (Color online) The log-log plot of the magnetization $\left\vert
\left\langle\protect\sigma_{x}\right\rangle\right\vert$ as a function of
bias $\protect\epsilon $ of the spin-boson model with (right) and without
(left) the RWA for $s=0.3$ (upper panel) and $s=0.7$ (lower panel). The
numerical results by VMPS are denoted by black circles, and the power-law
fitting curves are denoted  by the red dashed lines. All insets show the corresponding
linear plots. $\Delta =0.1$,$\protect\omega _{c}=1$, $\Lambda=2$, $L=50$, $%
d_{opt}=12$, and $D=20,40$ for $s=0.3,0.7$ respectively.}
\label{Fig6}
\end{figure}
Now we will study the nature of the second-order QPT in the sub-Ohmic
spin-boson model under RWA. The most important question is whether the RWA
changes the universality of the second-order QPT. The field relevant
order-parameter critical exponent $\delta $ can be determined through the
displayed power-law behavior $\left\langle \sigma _{x}\right\rangle \propto
\epsilon ^{1/\delta }$ at the critical coupling strength $\alpha =\alpha
_{c} $. Previously, various critical exponents in the full spin-boson model
have been calculated with different numerical approaches~\cite{QMC,
ED,Zhang, Chin, guo2012critical}. It is generally accepted that the exponent
$\delta $ takes the mean-field value $1/\delta =1/3$ for $s<1/2$, and the
non-classical one $1/\delta =\left( 1-s\right) /\left( 1+s\right) $ by the
exact hyperscaling for $s>1/2$ .

We present the magnetization by the VMPS method as a function of bias $%
\epsilon $ in a log-log plot for both $\lambda =1$ (RWA) and $\lambda =0$
(non-RWA) in Fig.~\ref{Fig6}. A very nice power-law behavior over three
decades is demonstrated in all cases. The results for the full model are
nearly the same as those in  Ref. \cite{guo2012critical}. It is
interesting to find that $1/\delta $ is around $1/3$ for $s<1/2$ for both
RWA and non-RWA cases. Very surprisingly, even for $s=0.7$ in the RWA model,
$1/\delta $ is still close to that in the full model within the statistical
errors, also indicating  hyperscaling. Our results suggest that
counterrotating terms would have no effect on critical exponent $\delta $
in the spin-boson model.

\subsection{Extensions to the Ohmic bath.}

\begin{figure}[tbp]
\centerline{\includegraphics[scale=0.4]{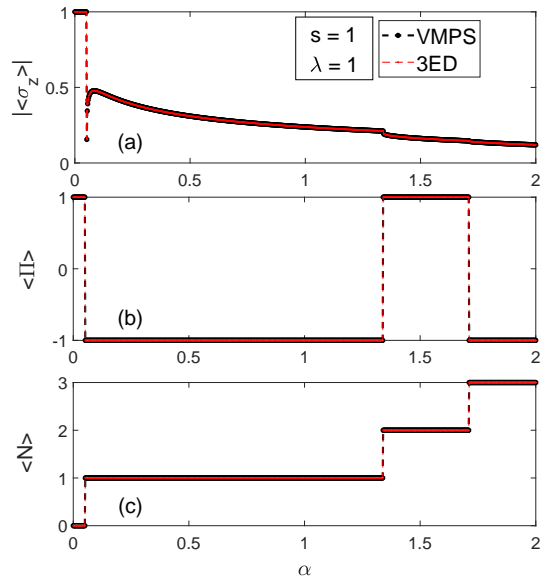}}
\caption{ (Color online) (a) The magnetization $\left\vert \left\langle%
\protect\sigma_{z}\right\rangle\right\vert$, (b) parity $\left\langle \hat{%
\Pi}\right\rangle$, and (c)  total excitation number $\left\langle N
\right\rangle$ as a function of the coupling strength within VMPS and $3$ED
approaches for the Ohmic bath. $s = 1$, $\protect\lambda =1$, $\Delta =0.1$,
$\protect\omega _{c}=1$, $\protect\epsilon =0$, $\Lambda=2$, $L=50$, $%
d_{opt}=12$, and $D=20$. }
\label{Ohmic}
\end{figure}

Now we turn to the Ohmic spin-boson model under the RWA. By both VMPS and
 $3$ED methods, the above observables are also calculated. In Fig.~\ref%
{Ohmic}, we collect magnetization $\left\langle \sigma _{z}\right\rangle $,
parity $\left\langle \hat{\Pi}\right\rangle $, and  total excitations $%
\left\langle N\right\rangle $ as a function of the coupling strength $\alpha
$ in the range $(0,2)$. During this wide regime, one can find that the model
undergoes a few first-order QPTs with the increment $1$ of the total
excitation number. The two approaches almost give the same results. It
follows that in the Ohmic bath, the $3$ED approach is suited to the wide
coupling regime where no second-order QPT occurs. This result is obviously
different from that in Ref. \cite{Tong} where only $N=1$, i.e., single
excitation, is considered. We have also studied the super-Ohmic bath $s=3/2$%
, and find that there is only one first-order QPT in the regime $\alpha
=[0,2]$ (not shown here).

\section{Conclusion}

In this paper, we study the spin-boson model in the RWA by the VMPS method,
MCS variational ansatz, and exact diagonalizations within the truncated
Hilbert space. Surprisingly, we find the second-order QPT in the RWA model
for any bath exponent $s<1$ for the first time, to the best of our
knowledge. A rich picture for the quantum phase transitions is observed.
Besides the second-order phase transition, the first-order phase transition
also appears in the same model, which could only vanish for small bath
exponents. The coexistence of both first- and second-order QPTs in the same
model has never been observed in  other spin-boson-like models, such as
the quantum Rabi and Dicke models in the RWA.

Within the statistical error, for all values of the bath exponents $s<1$,
the critical exponent $\delta$ is found to be nearly the same as those in
the full model. It is then suggested that the counterrotating terms would
have almost no effect on critical exponent $\delta$ in the critical regime.
The analytical argument about the quantum-to-classical mapping ~\cite%
{map,berry,Kirchner} in the spin-boson model in the RWA would be helpful to
account for this robust nature uniquely from the rotating-wave terms.

\textbf{ACKNOWLEDGEMENTS}  We acknowledge useful discussions with Stefan Kirchner. This work is supported by the National Science
Foundation of China (Nos. 11834005, 11674285), the National Key Research and
Development Program of China (No. 2017YFA0303002),

$^{*}$ Email:qhchen@zju.edu.cn

\setcounter{figure}{0}

\renewcommand{\thefigure}{A\arabic{figure}}

\begin{appendix}

\section{Convergence of results by the variational matrix product state
approach}

We provide  evidence for the full convergence of our VMPS results here.

\begin{figure}[tbp]
\centerline{\includegraphics[scale=0.4]{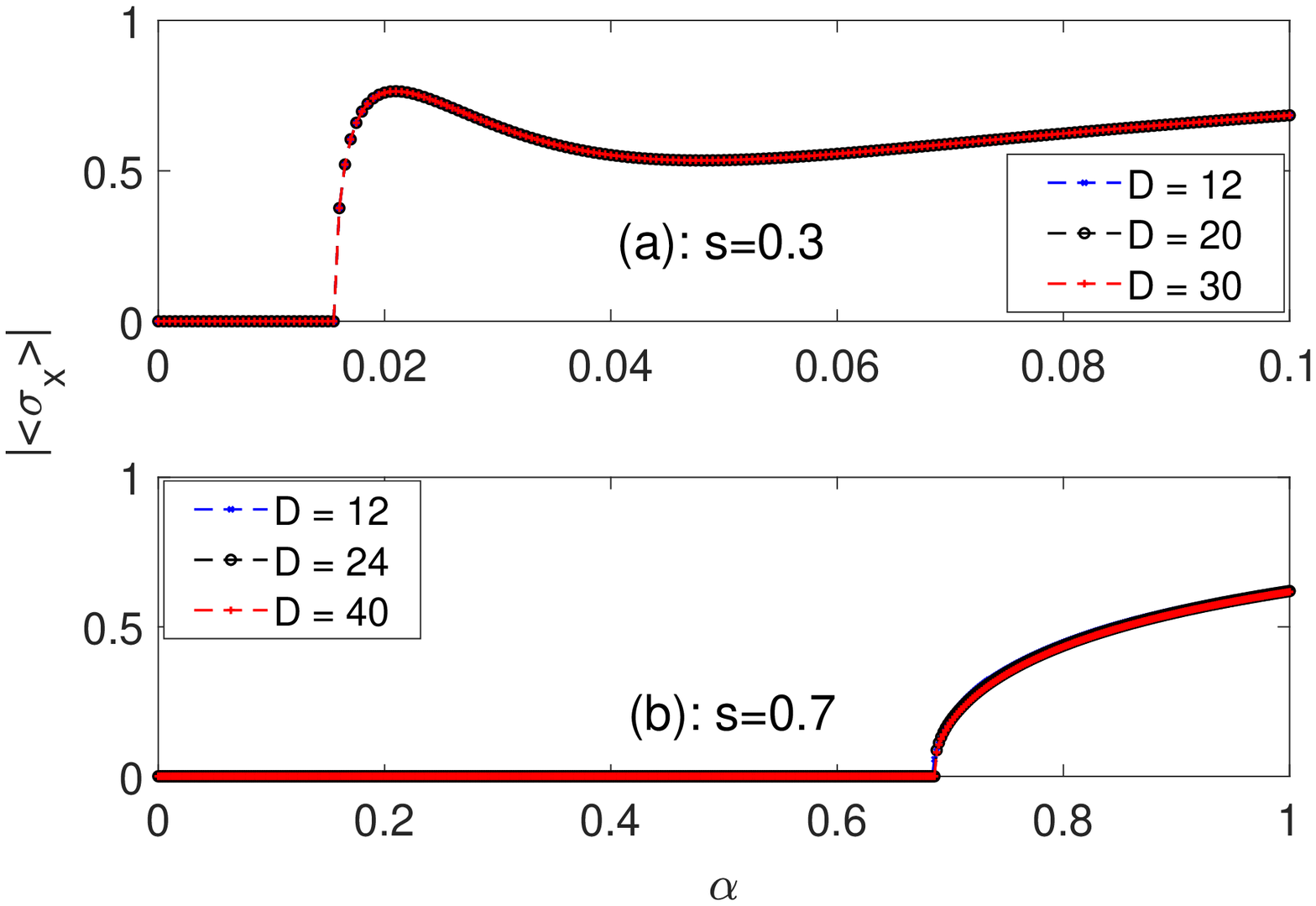}} %
\caption{ (Color online) The plot of the magnetization $\left\vert
\left\langle\protect\sigma_{x}\right\rangle\right\vert$ as a function of
coupling strength $\alpha$ with bond dimension (a) D=12, 20, 30 for $s=0.3$ and (b) $%
D = 12, 24, 40$ for $s=0.7$. Other parameters: $\protect\lambda =1$,
$\Delta =0.1$, $\protect\omega _{c}=1$, $\protect\epsilon = 0$,
$\Lambda = 2$, $L = 50$, $d_{opt} = 12$.}
\label{FigA4}
\end{figure}

The most important point of our paper is whether the second-order QPT occurs
in the spin-boson model in the RWA. To demonstrate this point convincingly,
we have  checked the magnetization as a function of the coupling
strength with increasing bond dimension (D). The results are shown in
Fig.~\ref{FigA4}. One can see that the magnetization is converged even for the smallest one $D=12$, indicating that the second-order QPT in our model is  reliable.

\begin{figure}[tbp]
\centerline{\includegraphics[scale=0.4]{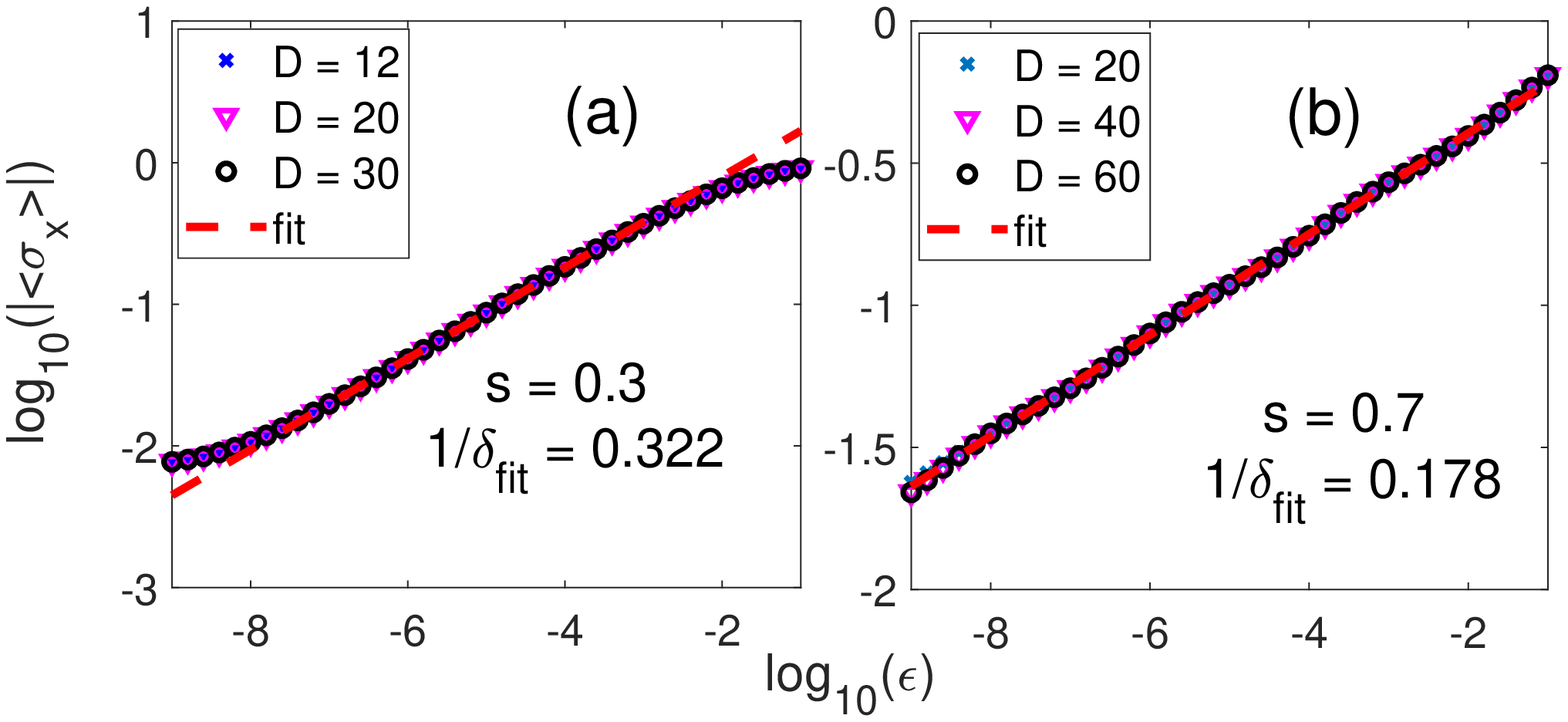}}
\caption{ (Color online) Convergence check for the VMPS parameter $D$ in the critical regime.
The panels shows the log-log plot of the magnetization $\left\vert
\left\langle\protect\sigma_{x}\right\rangle\right\vert$ as a function of
bias $\epsilon$ with bond dimension (a) $D=12, 20, 30$ for s=0.3 and (b)D = 20, 40, 60
for $s=0.7$. Other parameters: $\protect\lambda =1$, $\Delta =0.1$, $\protect%
\omega _{c}=1$, $\Lambda = 2$, $L = 50$, $d_{opt} = 12$.}
\label{FigA1}
\end{figure}

\begin{figure}[tbp]
\centerline{\includegraphics[scale=0.4]{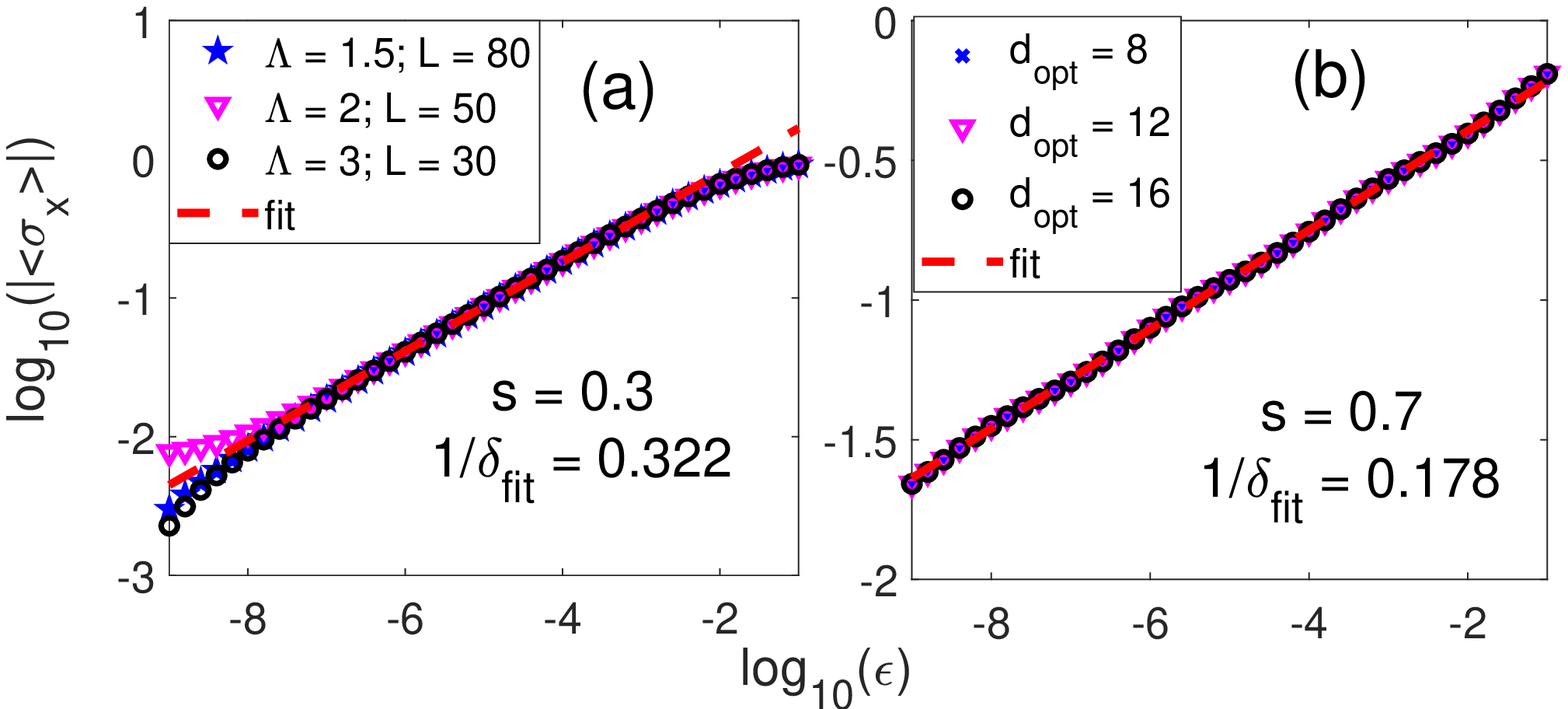}}
\caption{ (Color online) Convergence check for the VMPS parameter $\Lambda$ and $d_{opt}$ in the critical regime.
The panels show log-log plots of the magnetization $\left\vert
\left\langle\protect\sigma_{x}\right\rangle\right\vert$ as a function of
bias $\epsilon$ with (a)discrete logarithms $\Lambda = 1.5, L = 80$; $\Lambda = 2, L = 50$; and
$\Lambda = 3, L = 30$ for $s = 0.3$ with $D = 20$, $d_{opt} = 12$; and (b) optimized
physical dimension $d_{opt} = 8,12,16$ for $s = 0.7$ with $\Lambda = 2, L = 50$,
$D = 40$. Other parameters: $\protect\lambda =1$, $\Delta =0.1$, $\protect%
\omega _{c}=1$.}
\label{FigA2}
\end{figure}

The field relevant
order-parameter critical exponent $\delta $ should be determined precisely to study the universality class of the second-order QPT in this model. In doing so, we should  check its
convergence for the VMPS parameters carefully. Figures~\ref{FigA1} and ~\ref{FigA2} exhibit
the nice  power-law curves  with the same fitting exponent  for the selected bond dimension ($D$), optimized physical dimension ($d$),
and discrete logarithm ($\Lambda $) for the same values of the bath exponent $s$.  It is clearly demonstrated that we have obtained a convincing critical exponent  of the second-order QPT in this paper.

\begin{figure}[tbp]
\centerline{\includegraphics[scale=0.4]{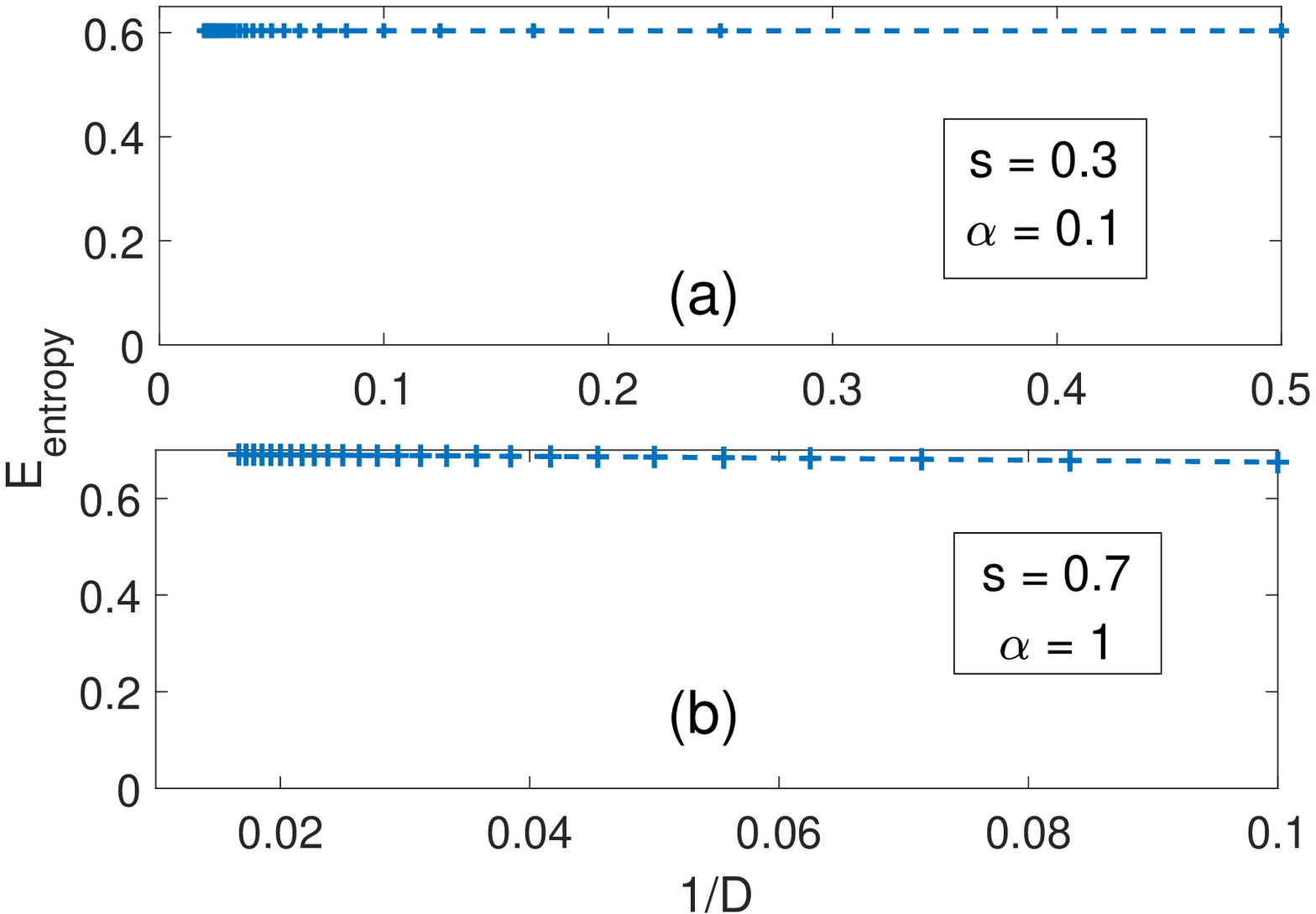}}
\caption{ (Color online) The plot of the entanglement entropy as a function
of $1/D$ (D is bond dimension) with the fixed coupling strength $\protect%
\alpha = 0.1,1$ for $s=0.3, 0.7$ respectively. Other
parameters: $\protect\lambda =1$, $\Delta =0.1$, $\protect\omega _{c}=1$, $%
\protect\epsilon = 0$, $\Lambda = 2, L = 50$, $d_{opt} = 12$ .}
\label{FigA5}
\end{figure}

Finally we turn to the convergence of the entanglement as a function of the
inverse of the bond dimension 1/D. In the spin-boson model, the entanglement
entropy can be rewritten as \cite{Chin}
\begin{equation}
E=-P_{+}\log _{2}P_{+}-P_{-}\log _{2}P_{-}
\end{equation}
where $P_{\pm }=\left( 1\pm \sqrt{\left\langle \sigma _{x}\right\rangle
^{2}+\left\langle \sigma _{y}\right\rangle ^{2}+\left\langle \sigma
_{z}\right\rangle ^{2}}\right) /2$, which describes the correlation between
the spin and the bosonic bath.

In Fig.~\ref{FigA5}, we plot the entanglement entropy as a function of 1/D
for s=0.3 (upper) and s=0.7 (lower) at fixed coupling strength. It is
clearly shown that the entropy  saturates even before  $D=20$ for both cases, providing
further strong evidence for the convergence of our results for $D \ge 20$, i.e.,
the value selected in our calculation. Thus we believe that our
algorithm has converged to the global minimum, the ground state is the
correct one, and the results given here should be reliable.

\end{appendix}


\end{document}